\documentclass[reprint,amsmath,amssymb,aps,floatfix,groupedaddress,superscriptaddress,nofootinbib,preprintnumbers]{revtex4-1}
\usepackage{float}
\usepackage{amsmath}
\usepackage[pdftex]{hyperref}
\usepackage{graphicx,color}
\usepackage{dcolumn}
\usepackage{bm}
\usepackage{here}
\usepackage{comment}
\usepackage[paperwidth=210mm,paperheight=297mm,centering,hmargin=1.5cm,vmargin=2.0cm]{geometry}
\usepackage{tabularx}

\allowdisplaybreaks[1]

\makeatletter
\newsavebox{\@brx}
\newcommand{\llangle}[1][]{\savebox{\@brx}{\(\m@th{#1\langle}\)}%
  \mathopen{\copy\@brx\mkern2mu\kern-0.9\wd\@brx\usebox{\@brx}}}
\newcommand{\rrangle}[1][]{\savebox{\@brx}{\(\m@th{#1\rangle}\)}%
  \mathclose{\copy\@brx\mkern2mu\kern-0.9\wd\@brx\usebox{\@brx}}}
\makeatother

\begin{document}

\title{Efficiency correction for cumulants of multiplicity distributions based on track-by-track efficiency} 

\author{Xiaofeng Luo}
\email{xfluo@mail.ccnu.edu.cn}
\affiliation{Key\,Laboratory\,of\,Quark\,\&\,Lepton\,Physics\,(MOE)\,and\,Institute\,of\,Particle\,Physics,\,Central\,China\,Normal\,University,\,Wuhan\,430079,\,China}
\author{Toshihiro Nonaka}
\email{tnonaka@rcf.rhic.bnl.gov}
\affiliation{Key\,Laboratory\,of\,Quark\,\&\,Lepton\,Physics\,(MOE)\,and\,Institute\,of\,Particle\,Physics,\,Central\,China\,Normal\,University,\,Wuhan\,430079,\,China}


\begin{abstract}
We propose a simplified procedure for the experimental application of 
the efficiency correction on higher order cumulants in heavy-ion collisions. By using the track-by-track efficiency, we can eliminate possible bias arising from the average efficiencies calculated within the arbitrary binning 
of the phase space. Furthermore, the corrected particle spectra is no longer necessary for the average efficiency estimation and 
the time cost for the calculation of bootstrap statistical error can be significantly reduced. 
\end{abstract}
\maketitle

\newcommand{\ave}[1]{\ensuremath{\langle#1\rangle} }

\section{Introduction}
\label{sec:intro}
Higher order cumulants of event-by-event conserved charge fluctuations 
are important observables to search for the 
QCD critical point~\cite{Gupta:2011wh,susceptibility,correlation,Stephanov:2011pb,Asakawa:2009aj,Luo:2017faz} in heavy-ion collisions.
In the beam energy scan (BES) program at RHIC, 
the STAR experiment has measured the cumulants up to the $4$th order 
of the net-charge and net-kaon multiplicity 
distributions~\cite{net_charge,Adamczyk:2017wsl},
and up to the $6$th order of the net-proton multiplicity 
distribution~\cite{net_proton,Luo:2015doi,NonakaQM}.
In the recent results of the net-proton 
multiplicity distribution, the non-monotonic behaviour 
of the fourth order fluctuation has been observed with 
respect to the collision energy~\cite{Luo:2015ewa}, 
which is quite similar to the theoretical prediction 
with the critical point~\cite{nonmonotonic}.  
Since there are still large uncertainties in low collision energies, 
the second phase of the Beam Energy Scan program (BES-II) will be carried out in 2019-2021 focusing 
on the collision energy of $\sqrt{s_{\rm NN}}=$7.7--19.6 GeV. 

One of the difficulties of measuring the higher order cumulants is the efficiency correction.
It is known that the value of cumulants are artificially changed from the true ones~\cite{eff_koch,tsukuba_eff_separate},
due to the fact that detectors miss some particles with the probability called efficiency. 
Some analytical formulas have been proposed to correct the measured cumulants 
with the assumption that the response function of the efficiency follows the binomial distribution 
~\cite{Bialas:1985jb,eff_kitazawa,eff_koch,eff_psd_volker,eff_xiaofeng,eff_psd_kitazawa,Nonaka:2017kko,Kitazawa:2017ljq}.
Recently, a few attempts to understand and correct for the possible non-binomial 
efficiencies have been discussed~\cite{binomial_breaking,NonakaQM,Nonaka:2018mgw}, 
but there are still large systematic uncertainties arising from 
how to determine the detector-response functions.
The efficiency correction with the binomial assumption is thus still important. 
Hereafter, let us call it "binomial correction" for simplicity. In the binomial correction, particles are counted separately 
in the "efficiency bin" where the efficiency changes, 
which are substituted into the correction formulas 
with the corresponding value of efficiencies.
Due to the huge calculation cost with large number of efficiency bins
by using the correction formulas based on 
the factorial moments~\cite{eff_koch,eff_psd_volker,eff_xiaofeng}, 
more efficient formulas have been proposed in which factorial 
cumulants are used in the derivation~\cite{eff_psd_kitazawa,Nonaka:2017kko,Kitazawa:2017ljq}.
Experimentally, single particle efficiencies can be computed 
by the MC detector simulations in terms of 
the various experimental observables like 
centrality, multiplicity, vertex position, 
transverse momentum, rapidity, azimuthal angle and so on.
Efficiency bins are then defined with respect to 
the track-wise variables in which the particles are counted. 
This binning is still arbitrary, which depends not only on the computing power 
for the detector simulations but also on 
the homogeneousness of the detector in the acceptance.
Therefore, additional systematic studies will be necessary 
on how many efficiency bins are enough. 
The most crucial thing is that we need to calculate the averaged 
efficiency at each efficiency bin with weighted by the true spectra. 
This indicates that the traditional efficiency correction based on average efficiency, can not be performed 
until the corrected spectra of identified particle is available. Fortunately, we found those difficulties can be overcome by using the so called track-by-track efficiency correction methods. 
By doing this, the corrected particle spectra is no longer needed. We can also reduce the potential systematic bias by using the average efficiency correction method and the calculation cost for bootstrap statistical errors calculations.

This paper is organized as follows. 
In Sec~\ref{sec:effcorr}, the efficiency correction with the binomial model 
is introduced. The correction formulas based on factorial cumulants will be shown as well.
In Sec.~\ref{sec:problem}, we discuss three difficulties 
of the efficiency correction in the experimental applications. 
We also use a numerical study to demonstrate one of the issues.
In Sec.~\ref{sec:solution}, a simple solution using the 
track-by-track efficiency is shown, followed by the toy model 
to check the validity of the new method.


\section{Efficiency correction}
\label{sec:effcorr}
\subsection{Cumulants and factorial cumulants}
The $m$-th order cumulant of the probability distribution function $P(N)$ 
is defined as 
\begin{eqnarray}
	\langle N^{m}\rangle_{\rm c} 
	&=& \frac{\partial^{m}}{\partial\theta^{m}}K(\theta)|_{\theta=0}, \\ 
	K(\theta) &=& \ln\sum_{N}e^{N\theta}P(N) = \ln\langle e^{N\theta}\rangle,
\end{eqnarray}
where $K(\theta)$ represents the cumulant generating function. 
Another quantity, factorial cumulants $\langle N^{m}\rangle_{\rm fc}$ 
are also defined as
\begin{eqnarray}
	\langle N^{m}\rangle_{\rm fc}
	&=& \frac{\partial^{m}}{\partial^{m}}K_{\rm f}(s)|_{s=1}, \\
	K_{\rm f}(s) &=& \ln\langle s^{N}\rangle,  
\end{eqnarray}
with $K_{\rm f}(s)$ being the factorial-cumulant generating function.
Cumulants and factorial cumulants are connected to each other, 
e.g, factorial cumulants are expressed in terms of cumulants: 
\begin{equation}
	\langle N^{m}\rangle_{\rm fc} = \langle N(N-1)\cdots(N-m+1)\rangle_{\rm c}.
\end{equation}

\subsection{Binomial model}
Let us assume that the probability distribution function $P(N)$ is 
observed as $\tilde{P}(n)$ through detectors. 
Some of $N$ generated particles are missed by the detectors, 
which leads to the observation of $n$ particles ($n\leq N$).
This finite probability to measure particles 
characterized by the detector is called efficiency.
When the efficiencies for generated particles are independent each other, 
the detection process can be described 
by the binomial distribution $B_{\varepsilon,N}(n)$:
\begin{eqnarray}
	\tilde{P}(n) &=& \sum_{N}P(N)B_{\varepsilon,N}(n), \\ 
	B_{\varepsilon,N}(n) &=& \frac{N!}{n!(N-n)!}\varepsilon^{n}(1-\varepsilon)^{N-n},
        \label{eq:P=PB}
\end{eqnarray}  
where $\varepsilon$ represents the efficiency.
It is known that in this situation the relationship between
factorial cumulants of $P(N)$ and $\tilde{P}(n)$ 
is given by~\cite{Nonaka:2017kko,Kitazawa:2017ljq}
\begin{eqnarray}
	\langle n^{m}\rangle_{\rm fc} 
	= \varepsilon^{m}\langle N^{m}\rangle_{\rm fc}.
	\label{eq:fc_bm}
\end{eqnarray}
Using Eq.~(\ref{eq:fc_bm}) and extending to the $M$ multi-variable case 
$P({\bm N})=P(N_{1},N_{2},...,N_{M})$,
the formulas up to the fourth order cumulant are shown below:
\begin{widetext}
\begin{eqnarray}
	\bigl<Q\bigr>_{\rm c} &=& \ave{q_{(1,1)}}_{\rm c}, 
        \label{eq:mk_1} \\ \nonumber \\
	\bigl<Q^{2}\bigr>_{\rm c} &=& \ave{q_{(1,1)}^{2}}_{\rm c} 
	+ \ave{q_{(2,1)}}_{\rm c} - \ave{q_{(2,2)}}_{\rm c}, 
        \label{eq:mk_2} \\ \nonumber \\
	\bigl<Q^{3}\bigr>_{\rm c} 
        &=& \ave{q_{(1,1)}^{3}}_{\rm c} 
        + 3\ave{q_{(1,1)}q_{(2,1)}}_{\rm c} - 3\ave{q_{(1,1)}q_{(2,2)}}_{\rm c}
        + \ave{q_{(3,1)}}_{\rm c} - 3\ave{q_{(3,2)}}_{\rm c}
        + 2\ave{q_{(3,3)}}_{\rm c}, \label{eq:mk_3}
        \\ \nonumber \\
	\bigl<Q^{4}\bigr>_{\rm c} 
        &=& \ave{q_{(1,1)}^{4}}_{\rm c} 
        + 6\ave{q_{(1,1)}^{2}q_{(2,1)}}_{\rm c} - 6\ave{q_{(1,1)}^{2}q_{(2,2)}}_{\rm c} 
	+ 4\ave{q_{(1,1)}q_{(3,1)}}_{\rm c} + 3\ave{q_{(2,1)}^{2}}_{\rm c}
	\nonumber \\
	&& + 3\ave{q_{(2,2)}^{2}}_{\rm c} - 12\ave{q_{(1,1)}q_{(3,2)}}_{\rm c} 
	+ 8\ave{q_{(1,1)}q_{(3,3)}}_{\rm c} -  6\ave{q_{(2,1)}q_{(2,2)}}_{\rm c}
        \nonumber \\
	&& + \ave{q_{(4,1)}}_{\rm c} - 7\ave{q_{(4,2)}}_{\rm c} 
        + 12\ave{q_{(4,3)}}_{\rm c} - 6\ave{q_{(4,4)}}_{\rm c}, 
	\label{eq:mk_4} 
\end{eqnarray}
with $q_{(r,s)}$ defined as
\begin{equation}
	q_{(r,s)} = 
        \sum_{i=1}^{M} (a_{i}^{r}/\varepsilon_{i}^{s}) n_{i}, 
        \label{eq:multi_q}
\end{equation}
\end{widetext}
where $M$ represents the number of efficiency bins, 
$n_{i}$ represents the number of particles, 
and $a_{i}$ represents the electric charge 
of particles in $i$th efficiency bin.
The concept of the efficiency bin started to be taken into account 
in Ref.~\cite{eff_psd_volker} inspired by experimental requirement, 
which will be discussed in the next section. 

\section{Difficulties in the Efficiency Correction}
\label{sec:problem}
In this section, we clarify the three difficulties in the 
experimental application of the efficiency correction. 
First, the details of the experimental procedures 
of the efficiency correction is explained to point 
out the difficulties. 
Second, a simple toy model is used to demonstrate one of those. 

\subsection{Experimental application}
Experimentally, single-particle efficiency can be determined
by the Monte-Carlo approach of the detector simulation 
with respect to track-wise variable like transverse momentum, 
rapidity and azimuthal angle. This kind of efficiency map can be computed precisely 
as long as the computing source permits.
Usually,  the efficiency map is divided into various efficiency bins based on the track-wise variables
and the averaged efficiency in each efficiency bin needs to be estimated by using the corrected spectra.
In the current analysis of the net-proton fluctuations 
at the STAR experiment, the proton identification method 
is different between low and high $p_{T}$ regions, 
which leads to the step-like dependence of the efficiency 
as a function of $p_{T}$~\cite{Luo:2015ewa}.
In this case, particles need to be counted separately 
for the two $p_{T}$ region in which the values of the efficiency 
are different, then the true cumulants 
in entire $p_{T}$ regions can be reconstructed based on factorial moments~\cite{eff_xiaofeng}.

If the number of efficiency bins $M$ is large, it is more efficient to apply the efficiency correction according to 
Eqs.~(\ref{eq:mk_1})--(\ref{eq:mk_4}), which is based on factorial cumulants. 
For the statistical errors estimation, 
the bootstrap method would be the realistic way. 
We can obtain a new distribution by random sampling 
from the original distribution with the same number of events to calculate cumulants.
This procedure is repeated with 100 times, 
and the standard deviation of the 100 cumulant values calculated from the new distributions
are taken as the statistical error. 
In order to take into account the correlation between different 
efficiency bins, the sampling would be performed based on 
the $M$ dimensional histogram.

Below are three main difficulties in the experimental implementation  
of the efficiency correction.
\begin{enumerate}
	\item We expected that it can
	provide more precise efficiency corrected cumulants when using the large number of efficiency bins.
	However, it is difficult to know how many efficiency bins are enough. 
	\item In order to obtain the averaged efficiency for each efficiency bin, we need to consider 
	the variation of the particle yields within the efficiency bin, which means the corrected spectra 
	is necessary. This is especially crucial issue for new data set in 
	new collision energy or collision system where the corrected 
	spectra is not available.
	\item The calculation cost on the bootstrap increases as $\propto n^{M}$, 
	which indicates that the statistical error estimation will be difficult 
	for many efficiency bins in view of the computing source. 
\end{enumerate}

\subsection{Toy model simulation}
\label{subsec:toymodel}
\begin{figure}[htbp]
\begin{center}
\includegraphics[width=170mm]{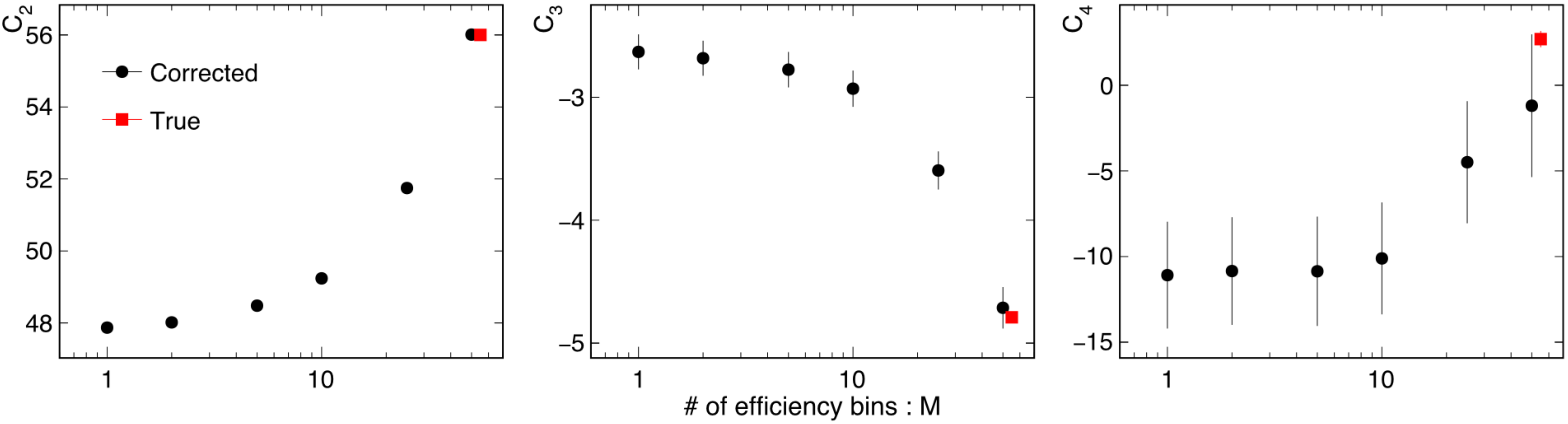}
\end{center}
\caption{ $C_{2}$, $C_{3}$ and $C_{4}$ as a function of efficiency bins. 
	The true value of cumulants are shown in red squares. }
\label{fig:conv_bin}
\end{figure}

Let us discuss more about the first issue related to the approximation 
of efficiency bins above by using a simple toy model.
We generated 50 independent binomial distributions 
for positively and negatively charged particles, 
$P^{\pm}_{i}(N^{\pm}_{i})$ $(i=1,2,...,50)$. 
The parameters of the binomial distributions are selected to be 
$N^{+}=4$, $N^{-}=3$ and $\varepsilon^{\pm}=0.8$ in Eq.~(\ref{eq:P=PB}). 
Efficiencies for each distribution ($\varepsilon^{\pm}_{i}$) are randomly chosen within   
$(0.5,1.0)$ with uniform distribution.
Particles $N^{\pm}_{i}$ are then randomly sampled with binomial efficiencies $\varepsilon^{\pm}_{i}$ 
to define the measured particles $n^{\pm}_{i}$.
Hereafter, let us suppose the net-particle distribution 
which consists of 50 independent distributions given by
$P_{\rm net}(N)$ with $N=\sum_{i=1}^{50}(N^{+}_{i}-N^{-}_{i}$).
We define the $m$-th order cumulant 
of $P_{\rm net}(N)$ as "true" cumulants.
The efficiency correction is performed with 
$M$ efficiency bins by using the corresponding averaged efficiency.  
For instance, let us suppose the efficiency correction with $M=10$ 
efficiency bins. In this case, whole 50 distributions are 
equally divided into $10$ sub-bins, 
each sub-bin contains $5$ distributions. 
The number of measured particles are counted at each sub-bin 
$n^{\pm}_{{\rm sub},x}$, $(x=1,2,...,10)$. 
Also the averaged efficiency at each sub-bin is given by 
\begin{equation}
\varepsilon^{\pm}_{{\rm sub},x}
	=\left\langle\sum_{i=5x}^{5(x+1)}\varepsilon^{\pm}_{i}N^{\pm}_{i}
	/\sum_{i=5x}^{5(x+1)}N^{\pm}_{i}\right\rangle,
\end{equation}
where the bracket represents the event average.
Then $n^{\pm}_{{\rm sub},x}$ and $\varepsilon^{\pm}_{{\rm sub},x}$ are substituted 
into Eqs.~(\ref{eq:mk_1})--(\ref{eq:multi_q}) 
event-by-event to calculate cumulants
\footnote{ Since the notation of electric charge is 
explicitly included in Eq.~(\ref{eq:multi_q}), 
we need following modification for substitution: 
$M\rightarrow2M$, $a_{i}=+1\;\;(i\leq M)$, $a_{i}=-1\;\;(i>M)$ }.
The efficiency correction has been done with different number of efficiency 
bins for $M=$1, 2, 5, 10, 25, 50 to check how many efficiency bins are needed 
to obtain the true cumulants.
Figure~\ref{fig:conv_bin} shows the corrected 
$C_{2}$, $C_{3}$ and $C_{4}$ as a function 
of the number of efficiency bins, 
where the true value of cumulants are shown in red squares. 
We find that the results with $M=50$ are consistent with the true cumulants, which is because 50 efficiency bins are assumed in the toy model.
It is also found that using wide efficiency bins is clearly incorrect and 
$M=$25 efficiency bins is still not enough. 

\onecolumngrid

\twocolumngrid

\section{Solution}
\label{sec:solution}
\subsection{Track-by-track efficiency}

Let us suppose infinite number of efficiency bins $M\rightarrow\infty$. 
Equation~(\ref{eq:multi_q}) is then written by

\begin{eqnarray}
	q_{(r,s)} &=& \sum_{i=1}^{\infty} (a_{i}^{r}/\varepsilon_{i}^{s}) n_{i} \\ 
	&=& \frac{a_{1}^{r}}{\varepsilon_{1}^{s}}n_{1} 
	+ \frac{a_{2}^{r}}{\varepsilon_{2}^{s}}n_{2} 
	+ \cdots
	+ \frac{a_{i}^{r}}{\varepsilon_{i}^{s}}n_{i} 
	+ \cdots \label{eq:q_trk_sum},
\end{eqnarray}
Since we consider that the width of the efficiency bin is now zero, 
each efficiency bin contains up to one particle. 
Thus, the summations for $n_{i}=0$ vanish 
(in other words, efficiency bins containing no particles 
don't need to be taken into account), 
and we immediately find that Eq.~(\ref{eq:q_trk_sum}) 
is equivalent to the summation with respect to the 
total number of particles $n_{\rm tot}$ in one event, 
which is given by
\begin{equation}
	q_{(r,s)} = \sum_{j=1}^{n_{\rm tot}} 
	\frac{a_{j}^{r}}{\varepsilon_{j}^{s}}, \label{eq:q_trk}
\end{equation}
which is connected with Eq.~(\ref{eq:multi_q}) 
via $n_{\rm tot} = \sum_{i=1}^{M} n_{i}$.
It is found that no variable related to the 
efficiency bin appears in Eq.~(\ref{eq:q_trk}). 
What we need to take care of is 
only the track-by-track efficiency, 
which indicates that 
the analytical formula of the efficiency 
with respect to track-wise variables 
can be directly used to determine 
the track-by-track efficiency 
\footnote{Dependence of the efficiency 
on event-wise variable 
can be also included if necessary.}.
Accordingly, we don't need to estimate 
the averaged efficiency at each efficiency bin, 
so the corrected spectra is no longer 
necessary for the efficiency correction.
The first two difficulties have been solved. 
Hereafter, let us call the correction 
in Eq.~(\ref{eq:multi_q}) "bin-by-bin" method, 
and call the track-wise correction in Eq.~(\ref{eq:q_trk}) 
"track-by-track" method.

\subsection{Method validation}
\begin{figure}[htbp]
\begin{center}
\includegraphics[width=160mm]{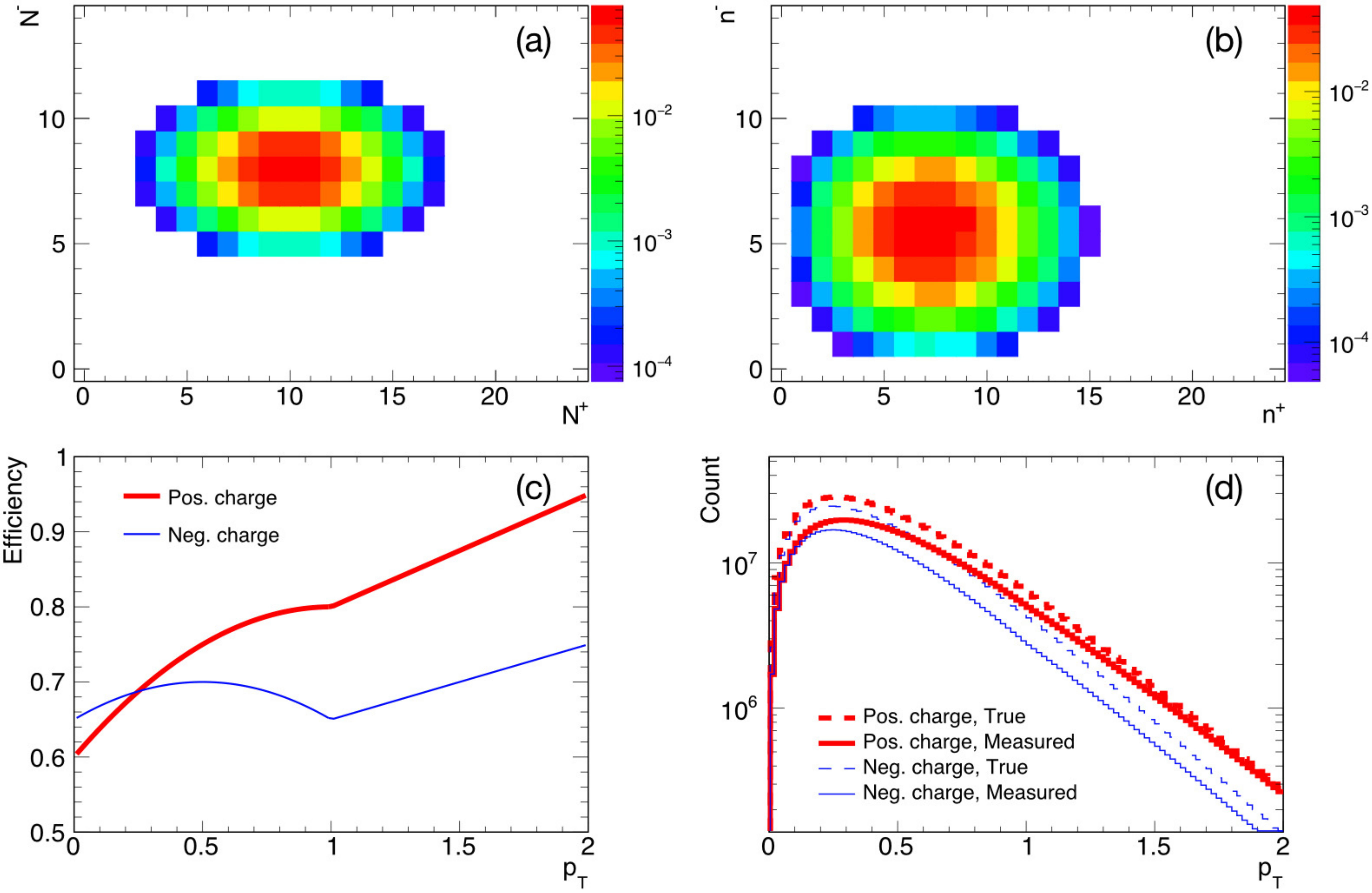}
\end{center}
\caption{ Correlation of the positively and negatively 
charged particles for the (a) true number distribution 
and (b) measured number distribution, 
(c) efficiencies as a function of $p_{T}$ and 
(d) true and measured $p_{T}$ spectra. }
\label{fig:pteff}
\end{figure}

In this sub-section, we employ a toy model 
to demonstrate the validity of the track-by-track efficiency method, 
and also discuss the rest one problem regarding 
how to estimate the statistical errors.
We start from two Gauss distributions, 
one is for positively charged particles, 
and the other is for negatively charged particles.
Figure~\ref{fig:pteff}--(a) shows the even-by-event correlation 
histogram between positively $N^{+}$ and negatively $N^{-}$ 
charged particles. 
For each particle $p_{T}$ is allocated according to the 
spectra given by
\begin{equation}
	f(p_{T})=p_{T}e^{-p_{T}/t}\;\;(0<p_{T}<2), 
\end{equation}
where $t=0.26$ and $0.24$ for positively and negatively charged 
particles, respectively.
$p_{T}$ dependent efficiency is assumed to be the convolution 
of the $1$st and $2$nd polynomial functions 
given by
\begin{eqnarray}
	\varepsilon(p_{T}) = \begin{cases} a \times p_{T}^{2} + b\times p_{T} + c\;\; (0<p_{T}<1),&\\ 
	d\times p_{T} + e\;\; (1<p_{T}<2),& \end{cases} \label{eq:effpt} 
\end{eqnarray}
where $(a,b,c,d)=$$(-0.2,0.4,0.6,0.15,0.65)$ and $(-0.2,0.2,0.65,0.1,0.55)$ 
for positively and negatively charged particles, respectively.
Each particle is then sampled by the corresponding value of efficiency 
determined in Eq.~(\ref{eq:effpt}). 
The resulting correlation between measured positively $n^{+}$ and negatively 
$n^{-}$ charged particles is shown in Fig.~\ref{fig:pteff}--(b), 
and $p_{T}$ spectra is shown in Fig.~\ref{fig:pteff}--(d).

Since efficiency changes continuously with respect to $p_{T}$, 
it is cumbersome to use the bin-by-bin correction 
formulas in Eq.~\ref{eq:multi_q}. 
Instead, we substitute the value of efficiency for each measured particle
determined by Eq.~(\ref{eq:effpt}) into Eq.~\ref{eq:q_trk} 
to calculate $q_{(r,s)}$ at each event. 

\begin{figure*}[htbp]
\begin{center}
\includegraphics[width=160mm]{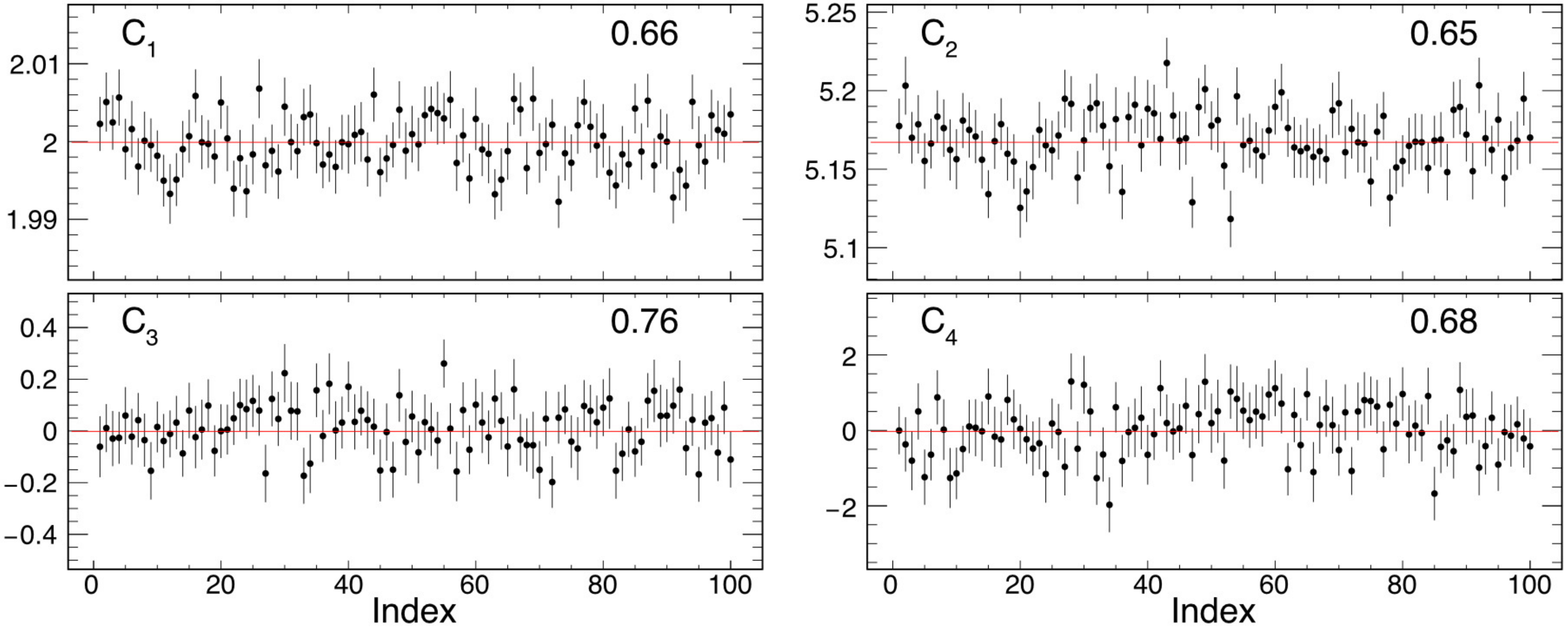}
\end{center}
\caption{ Efficiency corrected cumulants up to the $4$th 
order with statistical errors estimated by the bootstrap. 
Data points are calculated from 100 independent samples. 
Red lines represent the true value of cumulants. 
The number shown in right top of each panel shows 
the possibility of data points touching the true value out of 100 points.} 
\label{fig:bs}
\end{figure*}

The statistical errors can be estimated by the bootstrap method. 
As was mentioned in Sec.~\ref{sec:problem}, 
in the case of bin-by-bin correction method 
with $M$ efficiency bins, 
the bootstrap sampling would be performed based on 
the $M$ dimensional histogram.
But now there is no longer the efficiency bin, 
sampling can be simply done based 
on the spectra as follows:
\begin{enumerate}
	\item Resample $n_{1}$ and $n_{2}$ randomly from Fig.~\ref{fig:pteff}--(b)
		\label{enum:1} 
	\item Allocate $p_{T}$ for each particle based on the measured spectra in Fig.~\ref{fig:pteff}--(d)
		\label{enum:2} 
	\item Apply the efficiency correction to obtain efficiency corrected cumulants by using the known efficiency curve in Fig.~\ref{fig:pteff}--(c)
		\label{enum:3} 
	\item Repeat \ref{enum:1}--\ref{enum:3} with 100 times 
		and take the standard deviation as the statistical error. 
		\label{enum:4} 
\end{enumerate}
Above procedures are repeated with 100 times independently in order 
to check the validity of the correction and its statistical error. 
Results up to the fourth order are shown in Fig.~\ref{fig:bs}. 
It is found that the data points are distributed around the true value, 
so the correction method using track-by-track efficiency works well.
The probability of data points touching the true value 
within the statistical error is shown in the top right of each panel.
We find it comparable with the 1$\sigma$ nature of the Gaussian, 
which indicates the validity of the bootstrap.
We also checked that the calculation cost only depends 
on the number of particles $\propto n_{\rm tot}$, 
which is due to the fact that we don't have to 
consider particles which was not measured, 
while the bins containing no particles needed 
to be taken into account in bin-by-bin method.
On the other hand, we can also use the Delta theorem~\cite{Luo:2011tp,Luo:2013bmi} to calculate the 
statistical errors of the efficiency corrected cumulants, which will be implemented in the future data analysis. 

It should be noted that one would need to consider how to treat the momentum resolution in real 
experiments. Since we use $p_{T}$ of individual particles for the efficiency correction, the momentum resolution 
might directly affect the cumulants. This effect can be studied by smearing $p_{T}$ for each particle with the known value of 
momentum resolution. Finally, we note that bin-by-bin method (even single efficiency bin) 
using the averaged efficiency should also work in this toy model.
This is because only one probability distribution function 
is considered, which indicates we assume that 
the underlying physics is identical 
for whole $p_{T}$ region for each 
electric charge~\cite{Nonaka:2017kko,Kitazawa:2017ljq}.

\onecolumngrid

\twocolumngrid

\clearpage
\section{Summary}
\label{sec:summary}
In this paper, we pointed out three difficulties 
arising in the experimental application of the 
current efficiency correction method based on the 
arbitrary binning with respect to track-wise variables.
It was shown that those difficulties can be addressed by using the 
track-by-track efficiency in the efficiency correction formula 
based on factorial cumulants. 
We don't need to worry about how many efficiency bins are enough 
to calculate the efficiency corrected cumulants by using the analytical 
parameterization of the efficiency directly.
Thus, the averaged efficiency doesn't need to be estimated and the cumulant analysis can be proceeded without the corrected spectra.
Furthermore, the calculation cost for the statistical error estimation have been significantly reduced. 
Experimentally, single particle efficiency of the detectors 
would depend on transverse momentum, rapidity and azimuthal angle 
in one event. We assume that efficiencies for individual particles 
are independent, which leads to the binomial response function 
of the single particle efficiency.
One thing we have to work hard is to parametrize the single 
particle efficiency as a function of track-wise variables 
with the best precision as long as the computing source allows.
On the other hand, one should also study the possible non-binomial
efficiency effects in the experimental condition. Finally, we emphasize that the method shown in this paper could serve as 
one of the most precise and efficient way for the cumulant efficiency correction with binomial response function.
It will play an important role for the QCD critical point search and can be applied for the cumulant analysis in the future heavy-ion collision experiments, such as 
the BES-II program at RHIC, experiments at FAIR and NICA facilities.

\section{Acknowledgement}
This work is supported by the MoST of China 973-Project No. 2015CB856901, 
the National Natural Science Foundation of China under Grants 
(No. 11575069, 11828501, 11890711 and 11861131009), Fundamental Research Funds for the Central Universities No. CCNU19QN054 and China Postdoctoral 
Science Foundation funded project 2018M642878.

\bibliography{main}

\end{document}